\documentclass[a4paper,11pt]{article}
\pdfoutput=1 

\usepackage{jheppub} 

\usepackage[T1]{fontenc} 

\usepackage{pdfsync}
\usepackage{mathptmx}
\usepackage[utf8]{inputenc}
\usepackage[T1]{fontenc}
\usepackage{slashed}
\usepackage{times}
\usepackage{amssymb,amsfonts,amsmath,amsthm}
\usepackage{dsfont,bbm}
\usepackage{dcolumn}
\usepackage{epsf}
\usepackage{graphicx}
\usepackage[caption=false]{subfig}
\usepackage{dsfont}
\usepackage{simplewick}
\usepackage{bm}
\usepackage{eucal}
\usepackage{slashed}
\usepackage[active]{srcltx}
\usepackage[usenames]{color}

\title{\boldmath Non-integral geometry: additional term $f_A$ as a regularizing term
}

\author[a]{I.~V.~Anikin}
\affiliation[a]{Bogoliubov Laboratory of Theoretical Physics JINR, 141980 Dubna, Russia}

\emailAdd{anikin@theor.jinr.ru}

\abstract{
In the present paper, we first describe the principal basis of
non-integral geometry. Non-integral geometry
is a new field of generalized function (distribution) theory
where the effects breaking the symmetry of
integration measure have been investigated.
In turn, the non-symmetric integration measure
(the non-invariant measure)
leads to the complex form
of the universal, dimension-independent inverse operator
with the additional contributions compared to the methods of integral geometry.
The additional term with the complex integration measure
serves to the extension that improves the image reconstruction
procedure.
Then,
we prove that
 this additional term $f_A$
in the universal
inverse Radon transforms plays a role of the regularizing contribution.
In particular, we show that owing to the presence of $f_A$ the
corresponding complex singularities can be eliminated
in the image reconstruction process.
}

\begin{document}
\maketitle
\flushbottom

\section{Introduction}
\label{Intro}

Integral geometry as a theory studies the symmetry properties of the corresponding
measures on the geometrical spaces. In particular, the direct Radon transform (DRT)
considered as the singular delta-functional on the well-defined manifold is one
of the main objects in integral geometry. Besides, the integral measure
constructed by the differential forms in DRT possesses the certain symmetry
and leads to the invariant measure
\cite{Gelfand:1964, GGV}.

The methods inspired by integral geometry underlie
the mathematical tools for computerized tomography (CT), astrophysics and seismology
\cite{Deans}. Indeed, the inverse Radon transform (IRT)
are being used to reconstruct the image of an object under investigation.
It is worth to notice that
the inversion of Radon transforms meets the problems related to the ill-posedness.
It leads to the necessity to regularize the integral representation of the inverse operator
(the different methods of regularization can be found, for example, in  \cite{Anikin:2019oes, Anikin:2024vto}.

Recently, in  \cite{Anikin:2019oes, Anikin:2024vto,Anikin:2025uqv,Anikin:2025lta},
the regularization of inverse operator
has been implemented in the form that ensures the universality of inverse Radon transform.
The universality of inverse Radon transform means the independence of inversion
on the space dimension. In contrast to \cite{Anikin:2019oes, Anikin:2024vto,Anikin:2025uqv,Anikin:2025lta},
the old procedure of Radon transform inversion
has been realized in the odd and even dimension of space separately and independently each other.
The dependence of old inversion methods on the parity of space dimension can be traced from
the use of Courant-Hilbert's identities which have the different representations depending
on the space dimension \cite{Courant-Hilbert}.
Besides, in Courant-Hilbert's identities,
the angular integration has been always performed over the full region of
variations, {\it i.e.} in the full interval $(0,\, 2\pi)$.

According to \cite{Anikin:2019oes, Anikin:2024vto,Anikin:2025uqv,Anikin:2025lta},
the universal representation for the inverse Radon transform can
be derived with the help of the standard regularization procedure, used in the generalized function (distribution)
theory \cite{Gelfand:1964, GGV}, and without referring Courant-Hilbert's identities.

As demonstrated, if the original homogeneous outset function, which is under DRT,
is well-localized in the space and does not possess the symmetry properties,
the Radon $\varphi$-angular dependence receives the definite restrictions, instead of the full region $(0,\, 2\pi)$.
Owing to these angular restrictions, the universal IRT involves two contributions
$f_S$ and $f_A$, where the additional term $f_A$ is related to the complex integration measure.
This complexity is a very important founding and it opens a possibility to extend and to improve the Tikhonov regularization
needed for the different practical applications \cite{Anikin:2024vto}.
In addition, the inhomogeneity of the outset function support leads also to the existence of the
additional term $f_A$ independently of the angular restrictions.

All these above-mentioned arguments break the corresponding symmetry that underlies the standard
principles of integral geometry and, on the other hand, form the new principles of {\it non-integral geometry}.
Notice that for the practical applications, the proposed methods of non-integral geometry are more
suitable and adequate because the original objects
corresponding to the outset functions are naturally non-symmetric ones.

In the paper, we outline the basis of
non-integral geometry.
We demonstrate that non-integral geometry
as a new field of functional analysis
operates and studies
the effects arising from symmetry breaking of
the integration measure.
Moreover, the non-invariant (non-symmetric) measure
results in the complex form
of the universal and dimension-independent inversion
with the additional contributions.
It is important to notice that
the additional term with the complex integration measure
extends and improves substantively
the image reconstruction procedure.
According that, we prove that this additional term $f_A$
in the universal
inverse Radon transform plays a role of the regularizing contribution.
In particular, we show that owing to the presence of $f_A$ the
corresponding complex singularities can be eliminated
in the image reconstruction process \cite{AAS-Mod}.

\section{Non-integral geometry and universal form of inverse Radon transform}
\label{Sec-1}

\begin{figure}[t]
\centerline{\includegraphics[width=0.7\textwidth]{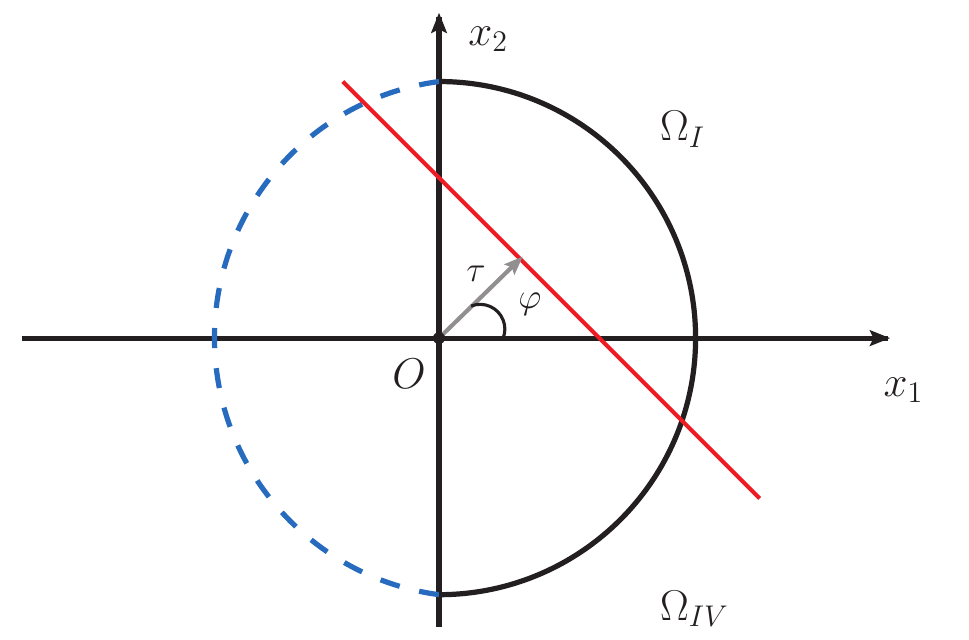}}
\caption{The restricted region in $\vec{\bf x}$-plane: $\Omega_I$ and $\Omega_{IV}$.
Notations: $\tau$ is the Radon radial variable, $\varphi$ is the Radon angular variable.
The line is parametrized as $\tau - \langle \vec{\bf n}_\varphi, \vec{\bf x} \rangle = 0$ with
$\vec{\bf n}_\varphi = (\cos\varphi, \, \sin\varphi)$.
For the sake of convenience, the starting point $O$ has been chosen to keep
the maximal possible symmetry.
}
\label{RT-1}
\end{figure}

To get started, we provide the basic definitions needed for our study.
In order to restore the internal structure
described by the given outset function $f(\vec{\bf x})$ ($\vec{\bf x}\in \mathbb{R}^n$ for $\forall n$), one has to use
the inverse (in our case, in the universal form) Radon transform (IRT).
As expected, the IRT-operation should express the outset function $f(\vec{\bf x})$ through the Radon image.
To obtain the representation of IRT, we first define
the direct Radon transform (DRT) of $f(\vec{\bf x})$ by (cf. \cite{Gelfand:1964, GGV})
\begin{eqnarray}
\label{F-t-4-dir}
G\equiv{\cal R}[f](\tau, \varphi, \theta_i) \quad \text{with} \quad
\mathcal{R}[f](\tau,\varphi, \theta_i)=
\int_{\mathbb{R}^n} d\mu_n( \vec{\bf x}) \, f(\vec{\bf x})\,
\delta\big( \tau - \langle\vec{\bf n}_{\varphi, \theta_i}, \vec{\bf x}\rangle\big),
\end{eqnarray}
where
$\vec{\bf n}_{\varphi, \theta_i}$ is the $n$-dimensional unit vector pointing along the radial Radon coordinate.
If so, the general form of IRT can be written as
\begin{eqnarray}
\label{Intro-1}
f ={\cal R}^{-1} G \quad \text{or} \quad
f(\vec{\bf x}) =
\int d\mu_n(\eta, \varphi, \theta_i)\,
{\cal R}[f]\big(\eta + \langle \vec{\bf n}_{\varphi, \theta_i}, \vec{\bf x}\rangle, \varphi, \theta_i\big),
\end{eqnarray}
where the integration measure depends now on the radial $\eta$ and the angular $\varphi, \theta_i$ Radon coordinates
(for the demonstration of Radon variables, see Fig.~\ref{RT-1}).

As shown, for example, in \cite{Anikin:2025lta},  the novel approach of Radon inversion
does not build upon Courant-Hilbert's identities which make the inversion to be substantially
different in the odd and even space dimensions.
Instead, we use the Fourier slice theorem that ensures the universality of inversion
\footnote{In this case, the universality of inversion means the independency of inversion
on the space dimension.}, see Appendix~\ref{App-02} for details.
However, the use of the Fourier slice theorem relates to the necessity of the corresponding regularization
owing to the hidden singularity encoded in the Fourier slice theorem, see Appendix~\ref{App-02}.
That is, in our approach,
the IRT-operator demands the regularization.
After the regularization procedure (see (\ref{int-lam}) of Appendix~\ref{App-02}), we obtain that
\begin{eqnarray}
\label{Inv-F-t-2-4}
f(\vec{\bf x}) \quad\Longrightarrow\quad f_\epsilon(\vec{\bf x})= {\cal R}_\epsilon^{-1} G
\quad \text{and}\quad
f_\epsilon(\vec{\bf x})
= f_S(\vec{\bf x}) + f_A(\vec{\bf x}),
\end{eqnarray}
where ${\cal R}_\epsilon^{-1}$ denotes symbolically the regularized IRT-operator,
while $f_S(\vec{\bf x})$ and $f_A(\vec{\bf x})$ take the following forms
\begin{eqnarray}
\label{Inv-F-t-2-4-S}
&&
\hspace{-0.5cm}
f_S(\vec{\bf x}) =
i^{\,n-2}(-)^{n-1} (n-1)!\, \int_{\Omega} d^{n-1}\mu(\varphi, \theta_i)\,
\int_{-\infty}^{+\infty} (d\eta)\, \frac{\mathcal{P}}{\eta^n}\,
\mathcal{R}[f](\eta + \langle \vec{\bf n}_{\varphi, \theta_i}, \vec{\bf x}\rangle, \varphi, \theta_i)
\end{eqnarray}
and
\footnote{To avoid the existence of surface terms, we here suppose $\mathcal{R}[f](\eta; ...)$ to be
a restricted function of $\eta$, see \cite{Anikin:2025lta}.}
\begin{eqnarray}
\label{Inv-F-t-2-4-A}
&&
\hspace{-0.8cm}
f_A(\vec{\bf x})= (-)^{n-1}
i^{\,n-1}\, \pi \int_{\Omega} d^{n-1}\mu(\varphi, \theta_i)\,
\int_{-\infty}^{+\infty} (d\eta)\,\delta(\eta)\,
\frac{\partial^{n-1}}{\partial\eta^{n-1}}\,\mathcal{R}[f](\eta + \langle \vec{\bf n}_{\varphi, \theta_i}, \vec{\bf x}\rangle, \varphi, \theta_i).
\end{eqnarray}
In these equations, $\mathcal{P}$ means the principle value as usual,
$\Omega$ defines the region of integration with respect to the angular Radon variables.
The exact domain of $\Omega$ has been specified below.

In the standard theory of integral geometry,
there are no the angular restrictions, {\it i.e.} we deal with the full regions of angular integrations.
As a rule, it is because of the certain symmetry of the outset function support, see below.
Moreover, if we believe additionally that the outset function $f(\vec{\bf x})$
is a homogeneous function with the trivial holonomy
(that is, no holonomy at all) then
$f_S$ of (\ref{Inv-F-t-2-4-S}) contributes only to the even dimension of space,
 while $f_A$ of (\ref{Inv-F-t-2-4-A}) contributes only to the
case of the odd dimension (cf. \cite{GGG}).
It can be readily demonstrated provided
$(a)$ we make the simultaneous replacements: $\eta\to -\eta$,
$\vec{\bf n}_{\varphi, \theta_i}\to - \vec{\bf n}_{\varphi, \theta_i}$
in (\ref{Inv-F-t-2-4-S}) and (\ref{Inv-F-t-2-4-A});
$(b)$ we take into account the corresponding symmetry
properties of $\eta$-dependent coefficient functions \cite{Deans}.
It is worth to notice that, in integral geometry, the term of (\ref{Inv-F-t-2-4-S}) contributes only to
the even-dimension space, while
the term of (\ref{Inv-F-t-2-4-A}) -- to the odd-dimension space. Moreover,
the contributing term is always real,
but the disappearing term is always imaginary.

However, in practice, the reconstructed image corresponding to the original outset function
is not symmetric and is inhomogeneous \cite{Anikin:2024vto,Anikin:2025uqv,Anikin:2025lta}.
The investigation of symmetry breaking effects together with the inhomogeneity of outset function
is a topic for a new branch of theory that can be called as {\it non-integral geometry}.
In this case, the angular integration measure receives the corresponding restrictions dictated by some asymmetry.
In addition, the regularized IRT-operator becomes the complex operator
(due to the fact that the disappearing, in the standard case, term now does contribute too)
which affects on the whole reconstruction procedure.

Thus, the sum of two contributions, see (\ref{Inv-F-t-2-4})-(\ref{Inv-F-t-2-4-A}), gives the
basis for the universal Radon inversion in the complex represention.

\section{Direct Radon transform of the probed outset function with the non-symmetric support
in $\mathbb{R}^2$}
\label{Sec-2}

In this section,
we present shortly the direct Radon transformation (DRT) on the chosen function $f(\vec{\bf x})$ with the simple support.

We begin with the outset function $f(\vec{\bf x})$, where $\vec{\bf x}\in \mathbb{R}^2$,
defined as (see Fig.~\ref{RT-1})
\begin{eqnarray}
\label{f-2}
f(\vec{\bf x}) = m_{I}\, \Theta\big( \vec{\bf x} \in \Omega_{I}\big) +
m_{IV}\, \Theta\big( \vec{\bf x} \in \Omega_{IV}\big).
\end{eqnarray}
In (\ref{f-2}) the characteristic functions
\footnote{The characteristic function is equal unit if the condition written as a argument takes place,
otherwise it is equal zero. It reminds the standard theta-function but without an ambiguity.}
 $\Theta$ are associated with
the supports given by  the following regions:
\begin{eqnarray}
\label{Omega-I}
&&\{\Omega_{I}: x_1>0, x_2>0, |\vec{\bf x}| \le 1 \}
\quad\text{and}
\\
\label{Omega-IV}
&&\{\Omega_{IV}: x_1 >0, x_2<0, |\vec{\bf x}| \le 1 \}.
\end{eqnarray}
The choice of two-dimensional space is mainly
dictated by CT-technology, but not only by simplicity. In practice, the different algorithms have been
designed for the two-dimensional reconstructions which work with the
corresponding transverse projections of the three-di\-men\-sional object,
see for example \cite{Anikin:2024vto,Anikin:2025uqv,Anikin:2025lta}.

By the direct calculation of Radon transformation (see Appendix~\ref{App-01} for details),
we obtain
that $\mathcal{R}[f](\tau, \varphi)$ contains the
nontrivial $\varphi$-angular dependence:
\begin{eqnarray}
\label{DRT-1}
\mathcal{R}[f](\tau, \varphi)&=&
M\, \sqrt{1-\tau^2}\,
\Theta_1\big( \tau,\,\varphi\big) +
M\,  \tau\, \cot\varphi\,
\Theta_2\big( \tau,\,\varphi\big) -
M\,  \tau\, \tan\varphi\,
\Theta_3\big( \tau,\,\varphi\big),
\end{eqnarray}
where
\begin{eqnarray}
\label{DRT-1-Theta}
&&
\Theta_1\big( \tau,\,\varphi\big)=
\Theta\big( \tau\in [\sin\varphi,\, 1]\big) +
\Theta\big( \tau\in [0,\, 1]\big) +
\Theta\big( \tau\in [\cos\varphi,\, 1]\big),
\nonumber\\
&&
\Theta_2\big( \tau,\,\varphi\big)=
\Theta\big( \tau\in [0,\,\sin\varphi]\big),
\quad
\Theta_3\big( \tau,\,\varphi\big)=
\Theta\big( \tau\in [\cos\varphi,\, 1]\big),
\end{eqnarray}
and $M=m_I+m_{IV}$.

Despite the Radon $\varphi$-angle varies in (\ref{DRT-1}) within the interval $(0,\,\pi/2)$,
the expression of (\ref{DRT-1}) involves both (\ref{Omega-I}) and (\ref{Omega-IV})
owing to the implemented replacements of variables.

It can be readily seen that if we extend the function support domain up to
the full disk
covering also $\Omega_{II}+\Omega_{III}$, in addition to (\ref{Omega-I}) and (\ref{Omega-IV}),
the Radon $\varphi$-angular dependence disappears provided $m_I=m_{IV}$.
In other words, in the case of the support defined by the symmetric disk
the angular dependence becomes to be degenerated.
It corresponds to the standard fact well-known from the literature.

\section{Universal form of Radon inversion in $\mathbb{R}^2$}
\label{Sec-3}

Using (\ref{Inv-F-t-2-4-S}) and (\ref{Inv-F-t-2-4-A}) for $\vec{\bf x}\in \mathbb{R}^2$,
we are going over to
the inverse Radon transformation (IRT) in the universal form.
As mentioned, the regularized and universal IRT is given by
\begin{eqnarray}
\label{IRT-1}
f_\epsilon(\vec{\bf x}) = f_S(\vec{\bf x}) + f_A(\vec{\bf x}),
\end{eqnarray}
where $f_S(\vec{\bf x})$ and $f_A(\vec{\bf x})$
are the standard and additional contributions, respectively, defined now as
\begin{eqnarray}
\label{IRT-fs}
&&f_S(\vec{\bf x})=-\int_0^{\pi/2} (d \varphi) \int_{-\infty}^{+\infty} (d\eta) \frac{{\mathcal{P}}}{\eta^2}
\mathcal{R}[f](\eta +
\langle\vec{\bf n}_\varphi, \vec{\bf x}\rangle, \varphi) ,
\\
\label{IRT-fa}
&&f_A(\vec{\bf x})=(-i\pi)\int_0^{\pi/2} (d \varphi) \int_{-\infty}^{+\infty} (d\eta) \delta(\eta) \frac{\partial}{\partial\eta}
\mathcal{R}[f](\eta +
\langle\vec{\bf n}_\varphi, \vec{\bf x}\rangle, \varphi).
\end{eqnarray}
In (\ref{IRT-fs}) and (\ref{IRT-fa}), the special attention should be paid
for the origin of complexity.
The complex prefactor
in (\ref{IRT-fa}) appears as a result of the corresponding analytical regularization of the inversion procedure,
(see \cite{Anikin:2024vto, Anikin:2025lta, Anikin:2025uqv}  in detail).
However, as one can see below, the regularization is not the only sources of complexity.

Inserting (\ref{DRT-1}) into (\ref{IRT-fs}) and (\ref{IRT-fa}), we get the following
representations:
\begin{eqnarray}
\label{IRT-fs-2}
&&f_S(\vec{\bf x})=f_S^{(\tau)}(\vec{\bf x}) +f_S^{(\varphi)}(\vec{\bf x}),
\\
\label{IRT-fa-2}
&&f_A(\vec{\bf x})=f_A^{(\tau)}(\vec{\bf x}) +f_A^{(\varphi)}(\vec{\bf x})
\end{eqnarray}
where
\begin{eqnarray}
\label{IRT-fs-3-1}
f_S^{(\tau)}(\vec{\bf x})&&=-M \int_0^{\pi/2} (d \varphi) \int_{-\infty}^{+\infty}(d\eta) \frac{{\mathcal{P}}}{\eta^2}
\sqrt{1-\big(\eta + \langle\vec{\bf n}_\varphi, \vec{\bf x}\rangle \big)^2}
\,
\Theta_1\big( \eta + \langle\vec{\bf n}_\varphi, \vec{\bf x}\rangle,\,\varphi\big),
\\
\label{IRT-fs-3-2}
f_S^{(\varphi)}(\vec{\bf x})&&=-M \int_0^{\pi/2} (d \varphi) \int_{-\infty}^{+\infty}(d\eta) \frac{{\mathcal{P}}}{\eta^2}
\Big( \eta + \langle\vec{\bf n}_\varphi, \vec{\bf x}\rangle \Big) \,
\nonumber\\
&&\times
\Big\{
\cot\varphi
\,
\Theta_2\big( \eta + \langle\vec{\bf n}_\varphi, \vec{\bf x}\rangle,\,\varphi\big)
-\,
\tan\varphi
\,
\Theta_3\big( \eta + \langle\vec{\bf n}_\varphi, \vec{\bf x}\rangle,\,\varphi\big)
\Big\}
\end{eqnarray}
and
\begin{eqnarray}
\label{IRT-fa-3-1}
\hspace{-0.6cm}
f_A^{(\tau)}(\vec{\bf x})&&\hspace{-0.2cm}=(-i\pi) M
\int_0^{\pi/2} (d \varphi) \int_{-\infty}^{+\infty} (d\eta) \delta(\eta) \frac{\partial}{\partial\eta}
\sqrt{1-\big(\eta + \langle\vec{\bf n}_\varphi, \vec{\bf x}\rangle \big)^2}
\,
\Theta_1\big( \eta + \langle\vec{\bf n}_\varphi, \vec{\bf x}\rangle,\,\varphi\big),
\\
\label{IRT-fa-3-2}
\hspace{-0.6cm}
f_A^{(\varphi)}(\vec{\bf x})&&\hspace{-0.2cm}= (-i\pi)M\int_0^{\pi/2} (d \varphi)
\int_{-\infty}^{+\infty} (d\eta) \delta(\eta) \frac{\partial}{\partial\eta}
\Big( \eta + \langle\vec{\bf n}_\varphi, \vec{\bf x}\rangle \Big) \,
\nonumber\\
&&\times
\Big\{
\cot\varphi
\,
\Theta_2\big( \eta + \langle\vec{\bf n}_\varphi, \vec{\bf x}\rangle,\,\varphi\big)
-\,
\tan\varphi
\,
\Theta_3\big( \eta + \langle\vec{\bf n}_\varphi, \vec{\bf x}\rangle,\,\varphi\big)
\Big\}.
\end{eqnarray}
In (\ref{IRT-fs-2}) and (\ref{IRT-fa-2}),
the $\eta$-integrations together with other computational reductions of
integrations are collected in Appendix.

It is important to notice that the restricted outset function support leads
to the nontrivial Radon $\varphi$-dependence
together with the different sources of complexity in the universal form of IRT.
Besides, as it is demonstrated by calculations,
the extended standard term of universal IRT involving both $f_S^{(\tau)}$ and $f_S^{(\varphi)}$
contains the analytical singularities which are and are not depending on the position $\vec{\bf x}$.
In this connection, as it turns out, the additional term $f_A$ plays the extremely
important role as a regularizer of these singularities, see below.

\section{The simplifications in the standard contributions}
\label{Sec-4}

Before going further, we consider the possible simplification in the standard contributions
which involve both the quadratic and linear $\tau$-dependence of DRT.
In particular,
we describe the cancellation which leads to the simplification of
the standard term.
As shown in Appendix~\ref{App-2}, for the tangent-part of the standard contributions,
it is convenient to split  $f_S^{(\varphi)}(\vec{\bf x})\big|_{\text{tan-term}}$
into two parts, see (\ref{IRT-fs-4-tan-2-App}).
Then, we have found that
the surface terms $J_{s.t.}$, which are traced from the standard contribution $f_S^{(\tau)}$,
see (\ref{Type-fs-2a-App}), are cancelled by some of terms from
$f_S^{(\varphi)}$, see
(\ref{IRT-fs-cot-b-App}) and (\ref{IRT-fs-tan-N2-b-App}). Indeed, we derive the following
relations:
\begin{eqnarray}
\label{Cancl-st-fS}
&&f_S^{(\tau)}(\vec{\bf x}) \Big|^{A_2}_{\text{surf.term}}
=
- M\, \int_0^{\pi/2} (d\varphi)
J_{s.t.}\big(A_{4};\, A_{2} \big)= - \,
f_S^{(\varphi)}(\vec{\bf x})\Big|^{(b)}_{\text{cot-term}},
\nonumber\\
&&
f_S^{(\tau)}(\vec{\bf x}) \Big|^{A_2^\star}_{\text{surf.term}}
=
- M\, \int_0^{\pi/2} (d\varphi)
 J_{s.t.}\big(A_{4};\, A_{2}^\star \big)= - \,
f_S^{(\varphi)}(\vec{\bf x})\Big|_{\text{tan-term}}^{N_2,\,(b)}.
\end{eqnarray}
Hence, we finally get that
\begin{eqnarray}
\label{Cancl-st-fS-2}
&&f_S^{(\tau)}(\vec{\bf x}) \Big|^{A_2}_{\text{surf.term}} +
f_S^{(\varphi)}(\vec{\bf x})\Big|^{(b)}_{\text{cot-term}}\equiv
- f_S^{(\varphi)}(\vec{\bf x})\Big|^{(b)}_{\text{cot-term}} +
f_S^{(\varphi)}(\vec{\bf x})\Big|^{(b)}_{\text{cot-term}}
=0,
\nonumber\\
&&f_S^{(\tau)}(\vec{\bf x}) \Big|^{A_2^\star}_{\text{surf.term}} +
f_S^{(\varphi)}(\vec{\bf x})\Big|_{\text{tan-term}}^{N_2,\,(b)}\equiv
- f_S^{(\varphi)}(\vec{\bf x})\Big|_{\text{tan-term}}^{N_2,\,(b)} +
f_S^{(\varphi)}(\vec{\bf x})\Big|_{\text{tan-term}}^{N_2,\,(b)}
=0.
\end{eqnarray}
Notice that, depending on the positions in $\Omega_I$ and $\Omega_{IV}$,
such terms as
$$f_S^{(\varphi)}(\vec{\bf x})\Big|^{(b)}_{\text{cot-term}}\quad
\text{and}\quad f_S^{(\varphi)}(\vec{\bf x})\Big|_{\text{tan-term}}^{N_2,\,(b)}$$
can produce
the $\vec{\bf x}$-depending singularities in $\Re{\rm e}$-space. In this sense, the
relations (\ref{Cancl-st-fS-2}) can be also treated as the regularizing conditions
leading to the corresponding singularity cancellations.

\section{The complexity of the standard contributions}
\label{Sec-5}

From (\ref{IRT-fs-2}) and (\ref{IRT-fa-2}), one can see that the contributions of $f_A$
are accompanied by the complex prefactors
\footnote{In general case, the complex prefactor can be associated with the
integration measure.}
 $(- i\pi)$ and there are no the other complexities
in $f_A$. So, in contrast to $f_S$, the term $f_A$ is the complex (imaginary) term by construction.
However, even the standard contributions of $f_S$ can also generate the definite
complexities owing to the cuts of the corresponding
logarithmic and polylogarithmic functions.

\subsection{$f_S^{(\tau)}$-term}
\label{Sec-5-1}

In (\ref{Type-fs-2d-App}), the $\varphi$-dependent  integrands:
$$J_{int.}^{log.}\big(A_{4};\, - A_{1} \big), \quad J_{int.}^{log.}\big(A_{4};\, A_{2} \big)
\quad\text{and}\quad
J_{int.}^{log.}\big(A_{4};\, A_{2}^\star \big)$$
in $f_S^{(\tau)}$ generate the complexities due to the
negative arguments of the corresponding logarithms:
$$\log\big(-A_1\big),\quad \theta\big(-A_2\big) \,\log\big(A_2\big)
\quad\text{and}\quad
\theta\big(-A_2^\star\big) \,\log\big(A_2^\star\big).$$
We thus have the following terms which correspond to the imaginary parts:
\begin{eqnarray}
\label{fStau-Im-1}
f_S^{(\tau)}(\vec{\bf x})\Big|_{log.}^{Im.}&&=
-\, M\, \int_0^{\pi/2} (d \varphi)
\frac{A_1(\varphi)}{\sqrt{1-A_1^2(\varphi)}}
\nonumber\\
&&
\times
\Big\{
i\, \alpha_1\, \pi \theta\big(  A_1(\varphi) \big) +
i\, \alpha_2\, \pi \theta\big(  - A_2(\varphi) \big) +
i\, \alpha_3\, \pi \theta\big(  - A_2^\star(\varphi) \big)
\Big\},
\end{eqnarray}
where the uncertainty parameters
\footnote{As well-known, $\log(-1)=i\pi (1+2k)$ where $k=0, \pm 1, \pm 2, ...$. For our
case, we choose two independent values $k=0$ and $k=-1$. It is reflected in
the uncertainty parameter $\alpha$.}
come from
$\log(-1)=i\, \alpha \pi$ with $\alpha=\pm 1$. In its turn, the theta-functions of (\ref{fStau-Im-1})
give the angular constraints as
\begin{eqnarray}
\label{phi-theta-1}
\theta\big(  A_1(\varphi) \big) &&\quad \Rightarrow \quad \cot\varphi > - \, \frac{x_2}{x_1},
\\
\label{phi-theta-2}
\theta\big( - A_2(\varphi) \big) &&\quad \Rightarrow \quad \cot\varphi >  \frac{1 - x_2}{x_1},
\\
\label{phi-theta-3}
\theta\big( - A_2^\star(\varphi) \big)&& \quad \Rightarrow \quad \cot\varphi >  \frac{1 - x_1}{x_2}.
\end{eqnarray}
In $\Omega_I$-region of $\vec{\bf x}$, the inequality (\ref{phi-theta-1}) is always true and, therefore,
there is no the influence on $\varphi$-integration;
the inequalities (\ref{phi-theta-2}) and (\ref{phi-theta-3})
give the restrictions
\begin{eqnarray}
\varphi < \arctan \frac{1-x_2}{x_1} \quad  \text{and} \quad
\varphi > \arctan \frac{1-x_1}{x_2}, \quad\text{respectively.}
\end{eqnarray}
In $\Omega_{IV}$-region of $\vec{\bf x}$,
the inequalities (\ref{phi-theta-1}) and (\ref{phi-theta-2}) reduce to (with  $x_2= - \tilde x_2$)
\begin{eqnarray}
\varphi < \arctan \frac{\tilde x_2}{x_1} \quad\text{and}\quad
\varphi < \arctan \frac{1+ \tilde x_2}{x_1}, \quad\text{respectively};
\end{eqnarray}
while the inequality (\ref{phi-theta-3}) always takes place and there are no the other
restrictions in this region.

\subsection{$f_S^{(\varphi)}$-term}
\label{Sec-5-2}

We are now going over to the discussion of the different sources of complexities
in $f_S^{(\varphi)}$. To this end, we present $f_S^{(\varphi)}$ as a sum of the structure contributions
(see (\ref{fS-sum-1}) and other equations in Appendix~\ref{App-2}):
\begin{eqnarray}
\label{fS-sum-W1}
f_S^{(\varphi)}(\vec{\bf x})&&=
f_S^{(\varphi)}(\vec{\bf x})\Big|_{\text{cot-term}} +
f_S^{(\varphi)}(\vec{\bf x})\Big|_{\text{tan-term}}=
\Big\{f_S^{(\varphi)}(\vec{\bf x})\Big|_{\text{cot-term}}^{(a)}
+f_S^{(\varphi)}(\vec{\bf x})\Big|_{\text{cot-term}}^{(b)}
\Big\}
\nonumber\\
&&
+
\Big\{f_S^{(\varphi)}(\vec{\bf x})\Big|_{\text{tan-term}}^{N_1}+
\Big[
f_S^{(\varphi)}(\vec{\bf x})\Big|_{\text{tan-term}}^{N_2,(a)}
+f_S^{(\varphi)}(\vec{\bf x})\Big|_{\text{tan-term}}^{N_2,(b)}
\Big]
\Big\},
\end{eqnarray}
where all of the structure contributions are given
by (\ref{IRT-fs-cot-a-App})-(\ref{IRT-fs-tan-N2-b-App}).
Hence, after the simplifications considered in Sec.~\ref{Sec-4},
we leave with $f_S^{(\varphi)}(\vec{\bf x})$ which can be written as
\begin{eqnarray}
\label{fS-sum-W2}
f_S^{(\varphi)}(\vec{\bf x})=
f_S^{(\varphi)}(\vec{\bf x})\Big|_{\text{cot-term}}^{(a)}
+
\Big\{f_S^{(\varphi)}(\vec{\bf x})\Big|_{\text{tan-term}}^{N_1}+
f_S^{(\varphi)}(\vec{\bf x})\Big|_{\text{tan-term}}^{N_2,(a)}
\Big\}.
\end{eqnarray}
Since the boundary of $\Omega_I$ and $\Omega_{IV}$ describing by
the semi-circle: $\big\{ \vec{\bf x} \in \Omega_{I, IV} :\, \,|\vec{\bf x}|=1 \big\}$ has to be excluded from
the consideration \cite{Anikin:2019oes, Anikin:2024vto,Anikin:2025uqv,Anikin:2025lta},
the logarithmic function of $f_S^{(\varphi)}(\vec{\bf x})\big|_{\text{tan-term}}^{N_1}$
(see (\ref{IRT-fs-4-2b-App}))
has a negative argument that gives the corresponding complexity.
To see that, we can rewrite (\ref{IRT-fs-4-2b-App})) in the form of
\begin{eqnarray}
\label{fs-phi-N1-log-1}
f_S^{(\varphi)}(\vec{\bf x})\Big|_{\text{tan-term}}^{N_1}= M\, \int_0^{\pi/2} (d \varphi) \,\tan\varphi
\Big[ i\alpha\pi +
\log
\frac{1- A_1(\varphi)}
{ A_1(\varphi)} +
\frac{A_1(\varphi)}
{ A_1(\varphi) - 1}
- 1
\Big],
\end{eqnarray}
where the condition $A_1(\varphi)< 1$ has been used for the logarithmic argument.
Notice that it is one of the reasons for the imaginary terms in the standard contributions.
The other reason of the possible complexity in (\ref{fs-phi-N1-log-1}) is the direct calculation of
the integrations. Having used the standard trigonometric replacement of integration variables,
the result of calculations in  (\ref{fs-phi-N1-log-1}) has been expressed through the logarithmic functions
with the possible negative arguments as well.

The similar situation takes place with
the terms $f_S^{(\varphi)}(\vec{\bf x})\big|_{\text{cot-term}}^{(a)}$
and $f_S^{(\varphi)}(\vec{\bf x})\big|_{\text{tan-term}}^{N_2,(a)}$.
Again, using the trigonometric replacements in the integrals of
(\ref{IRT-fs-cot-a-App}) and (\ref{IRT-fs-tan-N2-a-App}) for calculations,
we can readily obtain the logarithmic together with
polylogarithmic functions which generate the imaginary parts
due to
the negative arguments of logarithms  and the large (greater than unit) argument of
polylogarithms.

After some algebra, collecting only the complex terms ,
we get the following representation:
\begin{eqnarray}
\label{fS-con-tan-a-1}
&&f_S^{(\varphi)}(\vec{\bf x})\Big|_{\text{cot-term}}^{(a)} +
f_S^{(\varphi)}(\vec{\bf x})\Big|_{\text{tan-term}}^{N_2,(a)} \stackrel{M}{\sim}
\nonumber\\
&&
\Big\{ i \alpha_1 \frac{\pi}{4} \log\big( 1+ A^2\big)  + i \alpha_2 \frac{\pi}{4} \log\big( 1+ B^2\big) -
i \alpha_3 \pi \log\big( B\big)
\Big\}
\nonumber\\
&&
 - \Big\{ \text{the analogous term with replacements:} \,A\to A^\star; \, B\to B^\star \Big\},
\end{eqnarray}
where
\begin{eqnarray}
\label{A-B-combis}
A=\frac{1-x_2}{x_1}, \quad B=\frac{x_2}{x_1}, \quad
A^\star=\frac{1-x_1}{x_2}, \quad B^\star=\frac{x_1}{x_2}.
\end{eqnarray}
It is important to stress that in (\ref{fS-con-tan-a-1})
we have the complex contributions which are $\vec{\bf x}$-dependent ones and they
can be singular in the definite positions.
The cancellations of the possible singularities have been discussed in the next section.

To conclude this section, we have demonstrated and discussed all sources of complexities
appearing in the standard contributions. Notice that, in contract to the additional contributions $f_A$,
the standard contributions $f_S$ deal with the real integration measure. However,
the complexities appear from the direct calculation of integrals in IRT of the given outset function, see
(\ref{f-2}). Such a property is not because of the special form of the chosen outset-function,
but it is rather the general property of any IRT in the case of the non-symmetric supports of outset functions.

\section{The cancellation of complex singularities: the regularizing role of $f_A$}
\label{Sec-6}

In the preceding section, we have discussed the origination of the complex terms in the
standard contributions $f_S$.
We remind that, by construction, the additional contributions $f_A$ operate with the imaginary integration measure
ensuring the considered complexity and there are no other sources the complexities in $f_A$.
It turned out, in the case of possible complex singularities, the additional term $f_A$ plays the
crucial role in the corresponding cancellations (or in the regularization).
In this section, our special attention is paid to
the cancellation of complex (imaginary) singularities.

It is convenient to begin the study of the regularization procedure
with (\ref{fStau-Im-1}) that represents the one of complexity sources and
is related to $f_S^{(\tau)}$.
The $\varphi$-integration is given by the typical integral, it reads
\begin{eqnarray}
\label{Typ-Int-fStau}
f_S^{(\tau)}(\vec{\bf x})\Big|_{log.}^{Im.} \sim
\int_{C_1}^{C_2} (d \varphi)
\frac{A_1(\varphi)}{\sqrt{1-A_1^2(\varphi)}}=
\text{Arcsinh\big(T\big)} \Big|_{C_d}^{C_u},
\end{eqnarray}
where the integration limits $C_1$ and $C_2$ have been defined by the theta-functions in (\ref{fStau-Im-1}) ,
and
\begin{eqnarray}
\label{Int-Lim-1}
C_d= \frac{r\sin(C_1-\phi)}{\sqrt{1-r^2}}, \quad
C_u= \frac{r\sin(C_2-\phi)}{\sqrt{1-r^2}}.
\end{eqnarray}
In (\ref{Typ-Int-fStau}) and (\ref{Int-Lim-1}), the polar system for $\vec{\bf x}=(r,\, \phi)$ has been used as an intermediate
step. Further, if the integration limits $C_1=0$ and $C_2=\pi/2$ (this is a result of one of theta-functions)
then we derive that
\begin{eqnarray}
\label{Typ-Int-fStau-2}
f_S^{(\tau)}(\vec{\bf x})\Big|_{log.}^{Im.} \sim
\text{Arcsinh}\frac{x_1}{\sqrt{1-|\vec{\bf x}|^2}} +
\text{Arcsinh}\frac{x_2}{\sqrt{1-|\vec{\bf x}|^2}}.
\end{eqnarray}
On the other hand, in general case, we have  that
\begin{eqnarray}
\label{Arcsh-Log-1}
\text{Arcsinh}\big(z\big) = \log\big(z+\sqrt{1+z^2} \big), \quad \text{with}\quad z\in \mathbb{C}.
\end{eqnarray}
Hence, for the points $(x_1,\ x_2)\in\Omega_{I, IV}$ in positions $(0,\,1)$ and  $(1,\,0)$,
we deal with the singularities like  $\log[\infty]$.
Fortunately, the additional term of $f_A^{(\tau)}\big|^{(a)}$ (see (\ref{IRT-fa-4-1-App})) gives the same
contribution as (\ref{fStau-Im-1}) but with the opposite common sign provided
\begin{eqnarray}
\label{Par-1}
\alpha_1=\alpha_2=\alpha_3= 1.
\end{eqnarray}
 So, we observe the full cancellation independently whether
 the singularities have been generated or not.
 It can be presented as
 \begin{eqnarray}
 \label{Canc-fS-fA-1}
 f_S^{(\tau)}(\vec{\bf x})\Big|_{log.}^{Im.} + f_A^{(\tau)}(\vec{\bf x})\Big|^{(a)}=0.
 \end{eqnarray}
This is the first type of cancellations.

The next stage is to consider
the logarithmic term of $f_S^{(\varphi)}(\vec{\bf x})\big|_{\text{tan-term}}^{N_1}$
that
leads to the $\vec{\bf x}$-in\-de\-pen\-dent logarithmic singularity
\footnote{Here and below, we assume that $\log[\infty]\equiv \lim_{t\to\infty}\log(t)$.}:
\begin{eqnarray}
\label{N1-log-sing-1}
f_S^{(\varphi)}(\vec{\bf x})\Big|_{\text{tan-term}}^{N_1,\log[\infty]} =
i\, \alpha\pi\, M\, \int_0^{\pi/2} (d \varphi) \,\tan\varphi \, \stackrel{M}{\sim}\,
i\, \alpha\pi\, \log[\infty].
\end{eqnarray}
This type of singularities is the most dangerous owing to the $\vec{\bf x}$-independence.
However, there are the other terms which can generate the similar singularities in
\begin{eqnarray}
\label{Cancel-fS-1}
f_S^{(\varphi)}(\vec{\bf x})\Big|_{\text{cot-term}}^{(a)}\quad
\text{and}\quad  f_S^{(\varphi)}(\vec{\bf x})\Big|_{\text{tan-term}}^{N_2,(a)}.
\end{eqnarray}
Besides, the singularities produced in
the terms of (\ref{Cancel-fS-1})
depend on the positions in $\vec{\bf x}$-space.
We call them as $\vec{\bf x}$-dependent singularities.
In particular, at the points giving by $(0,\, 1)$ and $(1,\,0)$, we have the following
combination:
\begin{eqnarray}
\label{x-dep-sing-s1}
&&f_S^{(\varphi)}(\vec{\bf x})\Big|_{\text{cot-term}}^{(a),\log[\infty]}+
f_S^{(\varphi)}(\vec{\bf x})\Big|_{\text{tan-term}}^{N_2,(a),\log[\infty]} \stackrel{M}{\sim}\,
\nonumber\\
&&
\Big\{
- i \tilde\alpha_1 \frac{\pi}{2} \log[\infty] +  i \tilde\alpha_2 \pi \log[\infty] +  i \tilde\alpha_3 \pi \log[\infty]
\Big\}_{\text{at $(0,\,1)$}}
\nonumber\\
&&
+
\Big\{
- i \tilde\beta_1 \pi \log[\infty] +  i \tilde\beta_2  \frac{\pi}{2} \log[\infty] -  i \tilde\beta_3 \pi \log[\infty]
\Big\}_{\text{at $(1,\,0)$}}
\end{eqnarray}
and
\begin{eqnarray}
\label{x-dep-sing-s2}
f_S^{(\varphi)}(\vec{\bf x})\Big|_{\text{tan-term}}^{N_1,\log[\infty], \text{$\vec{\bf x}$-dep.}}
\stackrel{M}{\sim}\,
\Big\{
- i \alpha_5 \pi \log[\infty]
\Big\}_{\text{at $(1,\,0)$}}.
\end{eqnarray}
Further, we dwell on the additional $f_A$-contributions which can generate the
singularities.
We find that the term $f_A^{(\tau)}(\vec{\bf x})\big|^{(b)}$
(see (\ref{IRT-fa-4-11-App}))
at the points $(0,\,1)$ and $(1,\,0)$
gives the following singularities:
\begin{eqnarray}
\label{x-dep-sing-a1}
f_A^{(\tau)}(\vec{\bf x})\Big|^{(b),\log[\infty]} \stackrel{M}{\sim}\,
\Big\{
- i  \frac{\pi}{2}\log[\infty]
\Big\}_{\text{at $(0,\,1)$}} +
 \Big\{
- i  \frac{\pi}{2}\log[\infty]
\Big\}_{\text{at $(1,\,0)$}} = - i \pi \log[\infty],
\end{eqnarray}
while the additional term $f_A^{(\varphi)}$
(see (\ref{fA-sum-1})-(\ref{IRT-fa-5-12-App}))
at the points $(0,\,1)$ and $(1,\,0)$
leads to the following singularities:
\begin{eqnarray}
\label{x-dep-sing-a2-1}
&&f_A^{(\varphi)}(\vec{\bf x})\Big|_{\text{cot-term}}^{\log[\infty]}=
\Big\{
 i  \frac{\pi}{2}\log[\infty]
\Big\}_{\text{at $(0,\,1)$}}
\end{eqnarray}
and
\begin{eqnarray}
\label{x-dep-sing-a2-1}
&&f_A^{(\varphi)}(\vec{\bf x})\Big|_{\text{tan-term}}^{\log[\infty]}=
\Big\{
 i \pi \log[\infty]
\Big\}_{\text{at $(0,\,1)$}} +
\Big\{
 i \pi \log[\infty]  +  i  \frac{\pi}{2}\log[\infty]
\Big\}_{\text{at $(1,\,0)$}}.
\end{eqnarray}
Summarizing all terms with the singularities presented by
(\ref{N1-log-sing-1})-(\ref{x-dep-sing-a2-1}), we obtain
that the cancellation of singularities takes place
provided
\begin{eqnarray}
\label{cancel-par-1}
&&\tilde\alpha_2=\tilde\beta_1=1,\quad
\tilde\alpha_3=\tilde\beta_3=1,
\nonumber\\
&&
\tilde\alpha_1=\tilde\beta_2=1, \quad
\alpha_5= - \alpha=1.
\end{eqnarray}
Notice that this set of parameters is not unique, we are able to arrange the other two set in addition:
\begin{eqnarray}
\label{cancel-par-2}
&&\tilde\alpha_2= - \tilde\beta_1= - 1,\quad
\tilde\alpha_3= - \tilde\beta_3=1,
\nonumber\\
&&
\tilde\alpha_1=\tilde\beta_2=1, \quad
\alpha_5= - \alpha=1
\end{eqnarray}
and
\begin{eqnarray}
\label{cancel-par-3}
&&\tilde\alpha_2= - \tilde\beta_1= 1,\quad
- \tilde\alpha_3= \tilde\beta_3=1,
\nonumber\\
&&
\tilde\alpha_1=\tilde\beta_2=1, \quad
\alpha_5= - \alpha=1.
\end{eqnarray}
All three sets of parameters (\ref{cancel-par-1})-(\ref{cancel-par-3})  are independent, and they can be
equivalently used.
Symbolically, the second type of cancellations can be presented as
 \begin{eqnarray}
 \label{Canc-fS-fA-2}
&&f_S^{(\varphi)}(\vec{\bf x})\Big|_{\text{tan-term}}^{N_1,\log[\infty]}  +
f_S^{(\varphi)}(\vec{\bf x})\Big|_{\text{tan-term}}^{N_1,\log[\infty], \text{$\vec{\bf x}$-dep.}}
+
\Big\{
f_S^{(\varphi)}(\vec{\bf x})\Big|_{\text{cot-term}}^{(a),\log[\infty]}+
f_S^{(\varphi)}(\vec{\bf x})\Big|_{\text{tan-term}}^{N_2,(a),\log[\infty]}
\Big\}
\nonumber\\
&&
 +
f_A^{(\tau)}(\vec{\bf x})\Big|^{(b),\log[\infty]} +
\Big\{
f_A^{(\varphi)}(\vec{\bf x})\Big|_{\text{cot-term}}^{\log[\infty]} +
f_A^{(\varphi)}(\vec{\bf x})\Big|_{\text{tan-term}}^{\log[\infty]}
\Big\}=0.
 \end{eqnarray}
Thus, two types of the singularity cancellations given
by (\ref{Canc-fS-fA-1}) and (\ref{Canc-fS-fA-2})
present the main result of the paper.
We can see that the additional contribution $f_A$ with the
complex integration measure plays the important role of the regularizing term.

\section{Conclusions}
\label{Cons}

In conclusion, we have presented the basis of
non-integral geometry which can be treated as a new field of functional analysis.
In order to clarify the main differences in comparison with the standard integral geometry,
it is worth to mention the following matching.
The standard methods of integral geometry are based on:
\begin{itemize}
\item Courant-Hilbert's identities traced from Green's
formulae. It gives the different representations of inversion for the spaces with even and odd dimensions;
\item the invariant (symmetric) integration measure that leads to the
full integration region: $\varphi\in (0,\, 2\pi)$.
\end{itemize}
On the other hand, the proposed methods of non-integral geometry
are built upon the following stages:
\begin{itemize}
\item exception  of Courant-Hilbert's identities (Green's formulae) from the consideration.
Instead, we use the Fourier slice theorem
 that demands
the  corresponding regularization of inversion.
As a result, we have derived the universal representation of inversion for the
spaces independently on even and odd dimensions;
\item using of the non-invariant (non-symmetric) integration
measure that leads to the
restricted integration region, for example, $\varphi\in (-\pi/2,\, \pi/2)$.
\end{itemize}

Inspired by non-integral geometry,
we have studied
the effects of symmetry breaking related to
the integration measure.
As demonstrated, the non-invariant (non-symmetric) measure
results directly in the complexity
of the universal inversion with the additional contributions.

In the presented paper, we have advocated a unique role of the additional term $f_A$. Namely,
the term $f_A$ is very important to regularize the
universal IRT of the outset function located in some domain.
Despite we have considered the simplest form of the outset function,
the discovered role of the additional term as the regularizing contribution
is the very general property even for the practical outset function.
We have emphasized that the mentioned regularization is a very important for
the image reconstruction procedure \cite{AAS-Mod}.

\section*{Acknowledgements}
We thank A.I.~Anikina, V.A.~Osipov, O.I.~Streltsova and N.A.~Tiurin for useful and illuminating discussions.

\section*{Conflict of interest}

The author declares no conflict of interest.

\section*{Data Availability Statement}
This manuscript has no associated data or the data will not be deposited.

\appendix
\renewcommand{\theequation}{\Alph{section}.\arabic{equation}}
\section*{Appendix}


In this Appendix we present the needed technical details of calculations and
the result of analytical calculations of
$f_S$ and $f_A$.

\section{Fourier slice theorem: axis singularity vs. unboundedness}
\label{App-02}

We are now in a position to discuss in detail the fundamental theorem called as
the {\it Fourier slice theorem}.
As demonstrated in a series of papers \cite{Anikin:2019oes, Anikin:2024vto,Anikin:2025uqv,Anikin:2025lta},
this theorem plays the principle role in the derivation of universal Radon inversion.

In particular, we demonstrate that the unboundedness of Radon transforms and the
axis singularities, well-known in physics, are related each other.
This fact demands the corresponding regularization in a sense of generalized function theory.

First, we consider the well-defined Fourier transform
\footnote{This transform involves the infinite integration limits. The case of the finite integration limits
related to the periodic function is beyond of our study due to the fact that they can be matched with the help of
the corresponding replacement.}
of some regular function $f(\vec{\bf x})$ with $\vec{\bf x} \in \mathbb{R}^2$:
\begin{eqnarray}
\label{App-F-tr-1}
\mathcal{F}[f](\vec{\bf q}) =
\int_{-\infty}^{+\infty} (d^2\vec{\bf x}) e^{-i \langle \vec{\bf q}, \vec{\bf x} \rangle} \, f(\vec{\bf x}).
\end{eqnarray}
In this representation, the Fourier coordinates $(q_1,\, q_2)$ and the Cartesian coordinates $(x_1,\, x_2)$
are independent. Notice that the inverse Fourier transforms are also well-defined.

Let us define the straightforward line which can be parametrized by two ways:
\begin{eqnarray}
\label{L-p-1}
&&t - \langle \vec{\bf q}, \vec{\bf x} \rangle \equiv t - q_1x_1 - q_2x_2 = 0,
\\
\label{L-p-2}
&&z - x_1 - \xi x_2 =0.
\end{eqnarray}
In (\ref{App-F-tr-1}),
the Fourier and Cartesian coordinates are still independent each other
despite the imposed condition (\ref{L-p-1}).
The parametrization of (\ref{L-p-1}) has three parameters: $\{t, q_1, q_2\}$, while the parametrization of (\ref{L-p-2})
-- two parameteres: $\{z, \xi\}$.
Notice that the line parametrization (\ref{L-p-1}) is redundant compared to the line parametrization (\ref{L-p-2}).
Indeed, any line in $\mathbb{R}^2$, which is not coinciding (or parallelizing) with the corresponding axes, has only two points
of axis interceptions:
\begin{eqnarray}
\label{Intercep-1}
  \begin{cases}
  (0,\, t/q_2)\,\,\, \text{and}\,\,\, (t/q_1,\,0)\,\,\,& \text{for (\ref{L-p-1})},\\
  (0,\, z/\xi)\,\,\, \text{and}\,\,\, (z,\,0)\,\,\, &\text{for (\ref{L-p-2})}.
  \end{cases}
\end{eqnarray}
Hence, it is enough to have only two external parameters to parametrize properly the straightforward line.

Inserting the line parametrization condition (\ref{L-p-1})  in the form of
\begin{eqnarray}
\label{Unit-rep}
1= \int_{-\infty}^{+\infty} (dt)\, \delta\big( t - q_1x_1 - q_2x_2 \big),
\end{eqnarray}
we get that
\begin{eqnarray}
\label{App-F-tr-2}
&&\mathcal{F}[f](q_1, q_2) =
\int_{-\infty}^{+\infty} (d^2\vec{\bf x}) e^{-i \langle \vec{\bf q}, \vec{\bf x} \rangle} \, f(x_1, x_2)
\Big\{
\int_{-\infty}^{+\infty} (dt)\, \delta\big( t - q_1x_1 - q_2x_2 \big)
\Big\}
\\
\label{App-F-tr-2-2}
&&=
\int_{-\infty}^{+\infty} (dt)\,  e^{-i t}
\Big\{
\int_{-\infty}^{+\infty} (d^2\vec{\bf x}) \, f(x_1, x_2)
 \delta\big( t - q_1x_1 - q_2x_2 \big)
\Big\}.
\end{eqnarray}
It is important to stress that in (\ref{App-F-tr-2-2}) the expression in the curly brackets
is not yet giving the direct Radon transformation of $f(x_1, x_2)$ because
it depends on three parameter: $\{t, q_1, q_2\}$
instead of needed two
\footnote{
This statement is known as the following theorem:
if $f$ is a function of $n$ independent variables then the Radon
transform of $f$, $\mathcal{R}[f]$, depends on $n$ independent variables too.
The Radon transform is a bijection and lives on
$\mathbb{R}^1\times \mathrm{S}^{n-1}$, see for example \cite{Deans}.}.

In order to get the direct Radon transformation in (\ref{App-F-tr-2}), we introduce one external condition: $\xi=q_2/q_1$.
Practically, it means that one singles out the line $q_2=\xi q_1$ in $(q_1, q_2)$-plane with a slope parameter
$\xi$.
We thus have the following:
\begin{eqnarray}
\label{Fq2-Fxi-1}
\mathcal{F}[f](q_1, q_2)\quad \xrightarrow{\xi=q_2/q_1} \quad
\mathcal{F}[f](q_1, \xi q_1),
\end{eqnarray}
where in the {\it r.h.s.} a new set of
two independent parameters: $\{q_1, \xi\}$ is presented.

Hence, the representation (\ref{App-F-tr-2}) is being changed on the following
\begin{eqnarray}
\label{App-F-tr-3}
&&\mathcal{F}[f](q_1, \xi \,q_1) =
\int_{-\infty}^{+\infty} (dt)\,  e^{-i t}
\Big\{
\frac{1}{|q_1|}\int_{-\infty}^{+\infty} (d^2\vec{\bf x}) \, f(x_1, x_2)
 \delta\big( t/q_1 - x_1 - \xi x_2 \big)
\Big\}
\\
&&=
\int_{-\infty}^{+\infty} (dz)\,  e^{-i z q_1}
\Big\{
\int_{-\infty}^{+\infty} (d^2\vec{\bf x}) \, f(x_1, x_2)
 \delta\big( z - x_1 - \xi x_2 \big)
\Big\}\equiv
\int_{-\infty}^{+\infty} (dz)\,  e^{-i z q_1} \, \mathcal{R}[f](z, \xi),
\nonumber
\end{eqnarray}
where the replacement: $z=t/q_1$ has been implemented and the direct Radon transform of $f(x_1, x_2)$,
denoted as $\mathcal{R}[f](z, \xi)$,  has been extracted.

Thus, the Fourier slice theorem states that
{\it if the Fourier coordinates $(q_1, q_2)$ transforms to a system with the new coordinates $(q_1, \xi)$ where
$\xi=q_2/q_1$ then the new Fourier image of $f(x_1, x_1)$ relates to the Radon image of $f(x_1, x_2)$ as
\begin{eqnarray}
\label{App-F-tr-3}
\mathcal{F}[f](q_1, \xi \,q_1)=
\int_{-\infty}^{+\infty} (dz)\,  e^{-i z q_1} \, \mathcal{R}[f](z, \xi),
\end{eqnarray}
where the coordinates $q_1$  and $z$ are Fourier-conjugated, while
the $\xi$-coordinate for $\mathcal{F}[f](q_1, \xi \,q_1)$ and $\mathcal{R}[f](z, \xi)$ remains to be identical.}

In other words, one may claim that the Fourier image of the outset function $f(x_1, x_2)$ calculated
along a line with the slope parameter $\xi$ is Fourier-conjugated with the Radon image of the same function $f(x_1, x_2)$
calculated along the set of lines with the same slope parameter $\xi$.
Moreover, thanks for the Fourier slice theorem, the Fourier angular restrictions coincide
with the Radon angular restrictions. This advantage has been successfully used
in \cite{Anikin:2019oes, Anikin:2024vto,Anikin:2025uqv,Anikin:2025lta}.

The system of $(q_1, \xi)\equiv (q_1, \xi q_1)$ reminds the polar system for $\vec{\bf q}$ but it is not identical to that.
Indeed, in $(q_1, \xi q_1)$-system, if one of coordinates is zero, $q_1=0$, then the other is nullified too, $q_2\equiv\xi q_1=0$,
provided $\xi$ is finite. This property of system takes also place for the polar system of $\vec{\bf q}$: if $q_1=\lambda \cos\varphi$
and $q_2=\lambda \sin\varphi$ then the nullification of both $q_1$  and $q_2$ has been ensured by the only requirement: $\lambda =0$.
The nullification of the radial polar coordinate $\lambda$ leads to the degeneration of $\varphi$-dependence.
This effect is called as the {\it axis singularity}.

On the other hand, if $q_1\to 0$, and therefore $q_2\to 0$, then $\{ z\to\infty, |t|\le M < \infty \}$.
Therefore, $\mathcal{R}[f](z, \xi)$ might be restricted with respect of $z$-variables to ensure the convergency of integral,
see (\ref{App-F-tr-3}). However, as well-known, the Radon image $\mathcal{R}[f](z, \xi)$ of any well-localized outset function $f(x_1, x_2)$
is unbounded. This contradiction shows us that the axial singularity at $\lambda\to 0$ (or $q_1\to 0$) and the unboundedness of Radon image
at $z\to\infty$ have the similar nature and
they demand the corresponding regularizations similarly.

Saying that, we go over to the polar system for $\vec{\bf q}$.
For this system,
using the inverse Fourier transform, we can write down that
\begin{eqnarray}
\label{Inv-F-t-2}
f(\vec{\bf x}) =
\int_{-\infty}^{+\infty} d^2 \vec{\bf q}\,  e^{+i\langle\vec{\bf q},\vec{\bf x}\rangle} \,\mathcal{F}[f](\vec{\bf q})
\Big|_{\vec{\bf q}=\lambda\vec{\bf n}_{\varphi}}=
\int_{0}^{+\infty} d\lambda \lambda\,  \int d\mu(\varphi)\,
e^{+i\lambda\langle \vec{\bf n}_{\varphi}, \vec{\bf x}\rangle}\,
\mathcal{F}[f](\lambda, \varphi).
\end{eqnarray}
Then, we use the Fourier slice theorem (\ref{App-F-tr-3}) to get the following
\footnote{$\epsilon$ as a subscript of $f$ denotes the $\epsilon$-regularization that
should be used in (\ref{Inv-F-t-2-3}).}
\begin{eqnarray}
\label{Inv-F-t-2-3}
\hspace{-0.5cm}f_\epsilon(\vec{\bf x}) =
\int d\mu(\varphi)\,
\int_{-\infty}^{+\infty} (d\eta)\,
\mathcal{R}[f](\eta + \langle \vec{\bf n}_{\varphi}, \vec{\bf x}\rangle, \varphi)
\int_{0}^{+\infty} d\lambda \lambda \, e^{-i\lambda\eta}\Big|_{\epsilon\text{-reg.}},
\end{eqnarray}
where $``\epsilon\text{-reg.}"$ denotes the necessary regularization (see below) and
the integration limits for $\varphi$-angular variable is irrelevant at this moment.
From (\ref{Inv-F-t-2-3}), one can see that the integration over $\lambda$ has been factorized
in the part which demands the corresponding regularization.
This is the analogous effect of axis (unboundedness) singularity described above, but it manifests
in terms of the Radon $\eta$-radial variable at $\eta \to 0$.
For the $\epsilon$-regularization,
we do $\eta\to\eta - i\epsilon$ which provides the analytical regularization \cite{Gelfand:1964}.
In the most general case, the $\lambda$-integration reads
\begin{eqnarray}
\label{int-lam}
&&i^{\,2-n}\int_{0}^{+\infty} d\lambda \lambda^{n-1} \, e^{-i\lambda(\eta-i\epsilon)}=
i\,\frac{\partial^{n-1}}{\partial\eta^{n-1}} \int_{0}^{+\infty} d\lambda \, e^{-i\lambda(\eta-i\epsilon)}
\nonumber\\
&&
=
 (-)^{n-1} (n-1)! \frac{\mathcal{P}}{\eta^n} + i\pi \frac{\partial^{n-1}}{\partial\eta^{n-1}}\delta(\eta).
\end{eqnarray}
One can see that due to (\ref{int-lam}), the inverse Radon representation of (\ref{Inv-F-t-2-3})
becomes to be well-defined.

\section{DRT of the probe outset function}
\label{App-01}

In this section, we present the details of the direct Radon transformation which acts on the
probe outset function given by (\ref{f-2}).

Inserting (\ref{f-2}) into (\ref{F-t-4-dir}), we obtain the following
(here, $\vec{\bf x}=(x_1,\, x_2)$):
\begin{eqnarray}
\label{DRT-Ap01-1}
&&\mathcal{R}[f](\tau, \varphi) =
\int_{\Omega_I + \Omega_{IV}} d^2 \vec{\bf x} \, f(\vec{\bf x})\,
\delta\big(\tau - x_1 \cos \varphi- x_2 \sin \varphi\big)
\\
\label{DRT-Ap01-2}
&&=
\int_{-\pi/2}^{\pi/2} d \phi\, \int_{0}^{1} dr \, r\,
 f(r\cos\phi, r\sin\phi)\,
 \frac{1}{|\cos (\phi - \varphi)|}
\delta\Big( \frac{\tau}{\cos (\phi - \varphi)}  - r \Big)
\\
\label{DRT-Ap01-3}
&&=
m_I\, \tau \,\int_{0}^{\pi/2} d \phi \,
\frac{\Theta\Big( \frac{\tau}{\cos(\phi-\varphi)}\in [0,\,1] \Big)}{\cos^2(\phi-\varphi)}
+
m_{IV}\, \tau \, \int_{0}^{\pi/2} d \tilde\phi \,
\frac{\Theta\Big( \frac{\tau}{\cos(\tilde\phi+\varphi)}\in [0,\,1] \Big)}{\cos^2(\tilde\phi+\varphi)},
\end{eqnarray}
where the polar system for $\vec{\bf x}$ has been used in (\ref{DRT-Ap01-2}) and
the replacement: $\phi=-\tilde\phi$ has been done in the second term, with $m_{IV}$,
 of (\ref{DRT-Ap01-3}).

It is worth to stress that, as well-known, the integration with delta-function results in the
set of conditions encoded in the corresponding characteristic $\Theta$-functions,
see (\ref{DRT-Ap01-3}).

The next stage is to disentangle all conditions presented by $\Theta$-functions.
For the sake of simplicity, we first assume that the external $\varphi$-variable belongs
to the interval $(0,\, \pi/2)$. The extension to the full region, $\varphi\in (-\pi/2, \pi/2)$
is trivial, see below.
So, we begin with the first term of (\ref{DRT-Ap01-3}) which is proportional to $m_I$.
We have
\begin{eqnarray}
\label{Theta-mI-1-a}
&&\tau \ge 0,
\\
\label{Theta-mI-1-b}
&&\cos(\phi-\varphi) \ge 0,
\\
\label{Theta-mI-1-c}
&&\tau \le \cos(\phi-\varphi).
\end{eqnarray}
The condition (\ref{Theta-mI-1-a}) is trivial, by construction, and it can be omitted in our further discussion
unless it may lead to misunderstanding.
The condition  (\ref{Theta-mI-1-b}) leads to the restrictions, applied to the
integration $\phi$-variable,
given by
\begin{eqnarray}
\label{Theta-mI-b}
\Big\{
I_b: \quad - \pi/2 + \varphi \le \phi \le \pi/2 + \varphi
\Big\}.
\end{eqnarray}
In addition, we remind that the integration limits of the considered term is $(0,\, \pi/2)$.
Hence, the condition (\ref{Theta-mI-1-b}) is simply reducing to the integration interval:
\begin{eqnarray}
\label{Theta-mI-b-2}
I_b \,\, \cap \,\, [0,\, \pi/2] = [0,\, \pi/2].
\end{eqnarray}
While the condition (\ref{Theta-mI-1-c}) gives the following restrictions:
\begin{eqnarray}
\label{Theta-mI-c}
\Big\{
I_c: \quad -\varphi_\tau + \varphi \le \phi \le \varphi_\tau + \varphi
\Big\} \quad \text{with} \quad \arccos \tau = \pm \varphi_\tau
\end{eqnarray}
and it requires more detailed consideration. Indeed, we can readily derive that
\begin{eqnarray}
\label{Theta-mI-c-2}
I_c \,\, \cap \,\, [0,\, \pi/2] = \{ \text{a set of conditions} \}.
\end{eqnarray}
In (\ref{Theta-mI-c-2}), a set of conditions involves the following restrictions on the $\phi$-integration limits
and the corresponding conditions for the external parameters $\tau$ and $\varphi$:
\begin{eqnarray}
\label{Condition-1}
&& \phi \in [0, \, \pi/2] \quad \text{if} \,\,\,
   \begin{cases}
   \varphi_\tau + \varphi \ge \pi/2\\
   -\varphi_\tau + \varphi \le 0
   \end{cases}
\Rightarrow
  \begin{cases}
   \tau \le \sin\varphi\\
    \tau \le \cos\varphi
   \end{cases}
\nonumber\\
&&
\Rightarrow
  \begin{cases}
    \text{if} \,\, \varphi \le \pi/4, &  \text{then} \,\, \tau \le \sin\varphi \le \cos\varphi \\
     \text{if}\,\,  \varphi \ge \pi/4, &  \text{then} \,\, \tau \le \cos\varphi \le \sin\varphi
   \end{cases},
\end{eqnarray}
and
\begin{eqnarray}
\label{Condition-2}
&& \phi \in [0, \, \varphi_\tau + \varphi] \quad \text{if} \,\,\,
   \begin{cases}
   \varphi_\tau + \varphi \le \pi/2\\
   -\varphi_\tau + \varphi \le 0
   \end{cases}
\Rightarrow
  \begin{cases}
   \tau \ge \sin\varphi\\
    \tau \le \cos\varphi
   \end{cases}
\nonumber\\
&&
\Rightarrow
  \big\{
    \sin\varphi \le \tau \le \cos\varphi,  \,\,\,  \text{iff} \,\,\, \varphi \le \pi/4
   \big\}
\end{eqnarray}
and
\begin{eqnarray}
\label{Condition-3}
&& \phi \in [-\varphi_\tau + \varphi, \, \varphi_\tau + \varphi] \quad \text{if} \,\,\,
   \begin{cases}
   \varphi_\tau + \varphi \le \pi/2\\
   -\varphi_\tau + \varphi \ge 0
   \end{cases}
\Rightarrow
  \begin{cases}
   \tau \ge \sin\varphi\\
    \tau \ge \cos\varphi
   \end{cases}
\nonumber\\
&&
\Rightarrow
  \begin{cases}
    \text{if} \,\, \varphi \le \pi/4, &  \text{then} \,\, \tau \ge \cos\varphi \ge \sin\varphi \\
     \text{if}\,\,  \varphi \ge \pi/4, &  \text{then} \,\, \tau \ge \sin\varphi \ge \cos\varphi
   \end{cases},
\end{eqnarray}
and
\begin{eqnarray}
\label{Condition-4}
&& \phi \in [-\varphi_\tau + \varphi, \, \pi/2] \quad \text{if} \,\,\,
   \begin{cases}
   \varphi_\tau + \varphi \ge \pi/2\\
   -\varphi_\tau + \varphi \ge 0
   \end{cases}
\Rightarrow
  \begin{cases}
   \tau \le \sin\varphi\\
    \tau \ge \cos\varphi
   \end{cases}
\nonumber\\
&&
\Rightarrow
  \big\{
    \cos\varphi \le \tau \le \sin\varphi,  \,\,\,  \text{iff} \,\,\, \varphi \ge \pi/4
   \big\}.
\end{eqnarray}
Further, we dwell on
the second term of (\ref{DRT-Ap01-3}) which is proportional to $m_{IV}$.
We write
\begin{eqnarray}
\label{Theta-mIV-1-a}
&&\tau \ge 0,
\\
\label{Theta-mIV-1-b}
&&\cos(\tilde\phi + \varphi) \ge 0,
\\
\label{Theta-mIV-1-c}
&&\tau \le \cos(\tilde\phi + \varphi).
\end{eqnarray}
As above, the condition (\ref{Theta-mIV-1-a}) is trivial and it can be omitted.
The condition (\ref{Theta-mIV-1-b}) corresponds to the restrictions
\begin{eqnarray}
\label{Theta-mIV-b}
\Big\{
J_b: \quad -\pi/2 - \varphi \le \tilde\phi \le \pi/2 - \varphi
\Big\}.
\end{eqnarray}
Taking into account that the integration limits of the considered term is also $[0,\, \pi/2]$,
the condition (\ref{Theta-mIV-1-b}) modifies the integration interval for $\tilde\phi$ as
\begin{eqnarray}
\label{Theta-mIV-b-2}
J_b \,\, \cap \,\, [0,\, \pi/2] = [0,\, \pi/2 - \varphi].
\end{eqnarray}
In contrast to the first term of (\ref{DRT-Ap01-3}), where the condition (\ref{Theta-mI-b})
does not produce any additional constraints, the condition (\ref{Theta-mIV-b}) does produce the
constraint, see (\ref{Theta-mIV-b-2}).

Going over to the condition (\ref{Theta-mIV-1-c}), one can see that
the following restrictions take place:
\begin{eqnarray}
\label{Theta-mIV-c}
\Big\{
J_c: \quad -\varphi_\tau - \varphi \le \tilde\phi \le \varphi_\tau - \varphi
\Big\}.
\end{eqnarray}
Moreover, these constraints are being interfered with (\ref{Theta-mIV-b-2}) giving
the interval for the $\tilde\phi$-variable:
\begin{eqnarray}
\label{Theta-mIV-c-2}
J_c \,\, \cap \,\, [0,\, \pi/2 - \varphi] = [0,\, \varphi_\tau - \varphi]
\end{eqnarray}
provided (\ref{Theta-mIV-1-a}) and
\begin{eqnarray}
\label{Condition-2tilde}
 -\varphi_\tau - \varphi \le 0 \quad
\Rightarrow
  \begin{cases}
   \tau \le \cos\varphi, & \,\,\, \text{if}\,\,\, |\varphi_\tau| \ge |\varphi| \\
    \tau \ge \cos\varphi, & \,\,\, \text{if}\,\,\, |\varphi_\tau| \le |\varphi|
   \end{cases}
\quad \Rightarrow
  \big\{
    0 \le \tau \le 1
   \big\}.
\end{eqnarray}
From this, one can see that the integration limits for $\tilde\phi$ given by (\ref{Theta-mIV-c-2})
correspond to the natural conditions (\ref{Condition-2tilde}) only.

The final stage is to summarize all above-considered contributions with the appropriate
conditions. After the corresponding integrations with respect to $\phi$- and $\tilde\phi$-variables
(see (\ref{DRT-Ap01-3})),
we derive that
\begin{eqnarray}
\label{DRT-Ap01-f-1}
&&\mathcal{R}[f](\tau, \varphi)\Big|^{\varphi \in (0,\, \pi/2)}
\nonumber\\
&&
=
\Big\{
\Theta(\varphi \le \frac{\pi}{4}) \Theta(\tau \le \sin\varphi \le \cos\varphi) +
\Theta(\varphi \ge \frac{\pi}{4}) \Theta(\tau \le \cos\varphi \le \sin\varphi)
\Big\} m_I \, \tau \big[ \cot\varphi + \tan\varphi \big]
\nonumber\\
&&
+
\Big\{
\Theta(\varphi \le \frac{\pi}{4}) \Theta(\sin\varphi \le \tau \le \cos\varphi) +
\Big\} m_I \, \tau \Big[ \frac{\sqrt{1-\tau^2}}{\tau} + \tan\varphi \Big]
\nonumber\\
&&
+
\Big\{
\Theta(\varphi \le \frac{\pi}{4}) \Theta(\tau \ge \cos\varphi \ge \sin\varphi) +
\Theta(\varphi \ge \frac{\pi}{4}) \Theta(\tau \ge \sin\varphi \ge \cos\varphi)
\Big\} m_I \, 2\, \sqrt{1-\tau^2}
\nonumber\\
&&
+
\Big\{
\Theta(\varphi \ge \frac{\pi}{4}) \Theta(\cos\varphi \le \tau \le \sin\varphi) +
\Big\} m_I \, \tau \Big[ \frac{\sqrt{1-\tau^2}}{\tau} + \cot\varphi \Big]
\nonumber\\
&&
+
\Big\{
\Theta(0 \le \tau \le 1) +
\Big\} m_{IV} \, \tau \Big[ \frac{\sqrt{1-\tau^2}}{\tau} - \tan\varphi \Big].
\end{eqnarray}
The inclusion of $\varphi \in (-\pi/2, \, 0)$ in our consideration corresponds
to the formal addition of the contribution (\ref{DRT-Ap01-f-1}) where the
replacements: $m_I \to m_{IV}$ and $m_{IV} \to m_{I}$ have been implemented.

Finally, after some algebra, we get
\begin{eqnarray}
\label{DRT-Ap01-f-2}
&&\mathcal{R}[f](\tau, \varphi) =
M\, \sqrt{1-\tau^2}
\Big\{
\Theta\big( \tau\in [\sin\varphi,\, 1]\big) +
\Theta\big( \tau\in [0,\, 1]\big) +
\Theta\big( \tau\in [\cos\varphi,\, 1]\big)
\Big\}
\nonumber\\
&&
+ M \, \tau\, \cot\varphi \Big\{
 \Theta\big( \tau\in [0,\,\sin\varphi]\big)
\Big\}
- M \, \tau\, \tan\varphi \Big\{
\Theta\big( \tau\in [\cos\varphi,\, 1]\big)
\Big\}
\end{eqnarray}
where $M=m_I+m_{IV}$.
This expression coincides with (\ref{DRT-1}).

\section{The standard contributions with the quadratic $\tau$-dependence of DRT}
\label{App-1}

The standard contributions with the quadratic $\tau$-dependence of DRT are
\begin{eqnarray}
\label{IRT-fs-4-1-App}
f_S^{(\tau)}(\vec{\bf x})&&=-\, M\, \int_0^{\pi/2} (d \varphi)
\Big\{
\int_{- A_1(\varphi)}^{A_4(\varphi)} (d\eta) \frac{ \sqrt{
1-\big(\eta + A_1(\varphi)\big)^2} }{\eta^2}
\nonumber\\
&&
+
\int_{A_2(\varphi)}^{A_4(\varphi)} (d\eta) \frac{ \sqrt{
1-\big(\eta + A_1(\varphi)\big)^2} }{\eta^2}
+
\int_{A_2^\star(\varphi)}^{A_4(\varphi)} (d\eta) \frac{ \sqrt{
1-\big(\eta + A_1(\varphi)\big)^2} }{\eta^2}
\Big\}
\end{eqnarray}
where the shortened notations have been introduced as
(here and below, $\vec{\bf x}$-dependence is hidden in all $A$-functions)
\begin{eqnarray}
\label{A-fun-App}
&&A_1(\varphi)\equiv \langle\vec{\bf n}_\varphi, \vec{\bf x}\rangle
=  x_2\, \sin\varphi +  x_1\, \cos\varphi,
\quad
A_2(\varphi)=(1-x_2)\, \sin\varphi - x_1\, \cos\varphi,
\nonumber\\
&&
A_2^\star(\varphi)=(1-x_1)\, \cos\varphi - x_2\, \sin\varphi,
\quad
A_4(\varphi)=1 - A_1(\varphi).
\end{eqnarray}

Thanks for the integration by part, the typical $\eta$-integration in (\ref{IRT-fs-4-1-App}) can be presented as
\begin{eqnarray}
\label{Type-fs-1-App}
J =
\int_{A_{d}(\varphi)}^{A_{u}(\varphi)} (d\eta) \frac{ \sqrt{
1-\big(\eta + A_1(\varphi)\big)^2} }{\eta^2}=
J_{s.t.}\big(A_{d};\, A_{u} \big) + J_{int.}\big(A_{d};\, A_{u} \big),
\end{eqnarray}
where the surface term $J_{s.t.}$ and the integration term $J_{int.}$ read
\begin{eqnarray}
\label{Type-fs-2a-App}
&&J_{s.t.}\big(A_{d};\, A_{u} \big)=
- \,\frac{ \sqrt{
1-\big(\eta + A_1(\varphi)\big)^2} }{\eta}\Big|_{A_{d}(\varphi)}^{A_{u}(\varphi)},
\\
\label{Type-fs-2b-App}
&&
J_{int.}\big(A_{d};\, A_{u} \big)=
-
\int_{A_{d}(\varphi)}^{A_{u}(\varphi)} (d\eta)
\frac{1}
{\sqrt{
1-\big(\eta + A_1(\varphi)\big)^2} }
\Big\{
1+\frac{A_1(\varphi)}{\eta}
\Big\}.
\end{eqnarray}
Having calculated $\eta$-integration, the first term of (\ref{Type-fs-2b-App})
gives $\arcsin$-function, while the second term results in $\log$-function.
It reads
\begin{eqnarray}
\label{Type-fs-2c-App}
&&J_{int.}\big(A_{d};\, A_{u} \big)=
- \arcsin\big(\eta+A_1(\varphi)\big) \Big|_{A_{d}(\varphi)}^{A_{u}(\varphi)}
+J_{int.}^{log.}\big(A_{d};\, A_{u} \big)
\end{eqnarray}
with
\begin{eqnarray}
\label{Type-fs-2d-App}
&&J_{int.}^{log.}\big(A_{d};\, A_{u} \big)=
\frac{A_1(\varphi)}{\sqrt{1-A_1^2(\varphi)}}
\nonumber\\
&&
\times
\log
\frac{2\big(1-A_1^2(\varphi)\big)-2\eta A_1(\varphi)
+2 \sqrt{\big(1-A_1^2(\varphi)\big)\big(1-(\eta + A_1(\varphi)\big)^2}}{\eta}
 \Big|_{A_{d}(\varphi)}^{A_{u}(\varphi)}
\end{eqnarray}

\section{The standard contributions with the linear $\tau$-dependence of DRT}
\label{App-2}

The standard contribution
with the linear $\tau$-dependence together with the angular $\varphi$-dependence of DRT
can be written in the form of a sum as
\begin{eqnarray}
\label{fS-sum-1}
f_S^{(\varphi)}(\vec{\bf x})=
f_S^{(\varphi)}(\vec{\bf x})\Big|_{\text{cot-term}} +
f_S^{(\varphi)}(\vec{\bf x})\Big|_{\text{tan-term}}
\end{eqnarray}
where
\begin{eqnarray}
\label{IRT-fs-4-cot-1-App}
f_S^{(\varphi)}(\vec{\bf x})\Big|_{\text{cot-term}}&&=-\, M\, \int_0^{\pi/2} (d \varphi) \,\cot\varphi
\int_{- A_1(\varphi)}^{A_2(\varphi)} (d\eta)
\frac{\eta + A_1(\varphi)}{\eta^2}
\end{eqnarray}
and
\begin{eqnarray}
\label{IRT-fs-4-tan-1-App}
f_S^{(\varphi)}(\vec{\bf x})\Big|_{\text{tan-term}}&&= M\, \int_0^{\pi/2} (d \varphi) \,\tan\varphi
\int_{A_2^\star(\varphi)}^{A_4(\varphi)} (d\eta)
\frac{\eta +A_1(\varphi)}{\eta^2}.
\end{eqnarray}
For the sake of convenience, we are splitting the contribution
$f_S^{(\varphi)}(\vec{\bf x})\Big|_{\text{tan-term}}$
into two terms where one of them resembles the contributions $f_S^{(\varphi)}(\vec{\bf x})\Big|_{\text{cot-term}}$
of (\ref{IRT-fs-4-cot-1-App}).
We have
\begin{eqnarray}
\label{IRT-fs-4-tan-2-App}
f_S^{(\varphi)}(\vec{\bf x})\Big|_{\text{tan-term}}=
f_S^{(\varphi)}(\vec{\bf x})\Big|_{\text{tan-term}}^{N_1} +
f_S^{(\varphi)}(\vec{\bf x})\Big|_{\text{tan-term}}^{N_2}
\end{eqnarray}
with
\begin{eqnarray}
\label{IRT-fs-4-tan-3a-App}
&&f_S^{(\varphi)}(\vec{\bf x})\Big|_{\text{tan-term}}^{N_1}=
M\, \int_0^{\pi/2} (d \varphi) \,\tan\varphi
\int_{-A_1(\varphi)}^{A_4(\varphi)} (d\eta)
\frac{\eta + A_1(\varphi)}{\eta^2}
\\
\label{IRT-fs-4-tan-3b-App}
&&f_S^{(\varphi)}(\vec{\bf x})\Big|_{\text{tan-term}}^{N_2}=
 - M\, \int_0^{\pi/2} (d \varphi) \,\tan\varphi
 \int_{-A_1(\varphi)}^{A_2^\star(\varphi)} (d\eta)
\frac{\eta + A_1(\varphi)}{\eta^2}.
\end{eqnarray}
From (\ref{IRT-fs-4-tan-3b-App}), one can see that
\begin{eqnarray}
\label{fS-phi-N2}
A_2^\star(\varphi) \Longrightarrow A_2(\varphi)\quad \text{and}\quad
f_S^{(\varphi)}(\vec{\bf x})\Big|_{\text{tan-term}}^{N_2}
\Longrightarrow
f_S^{(\varphi)}(\vec{\bf x})\Big|_{\text{cot-term}}
\end{eqnarray}
provided the replacements:
\begin{eqnarray}
\label{App-Rep-1}
x_1\,\leftrightarrow\, x_2, \quad \sin\varphi\,\leftrightarrow\,\cos\varphi.
\end{eqnarray}
At the same time, $A_1(\varphi)$ remains invariant under (\ref{App-Rep-1}).

After $\eta$-integration, we get the following expressions:
\begin{eqnarray}
\label{IRT-fs-4-2a-App}
f_S^{(\varphi)}(\vec{\bf x})\Big|_{\text{cot-term}}&&=-\, M\, \int_0^{\pi/2} (d \varphi) \,\cot\varphi
\nonumber\\
&&\times
\Big[
\log
\frac{x_1\,\cos\varphi - (1-x_2)\,\sin\varphi}
{ x_1\,\cos\varphi  + x_2\,\sin\varphi} +
\frac{x_1\,\cos\varphi + x_2\,\sin\varphi}
{ x_1\,\cos\varphi - (1-x_2)\,\sin\varphi}
- 1
\Big],
\end{eqnarray}
and
\begin{eqnarray}
\label{IRT-fs-4-2b-App}
f_S^{(\varphi)}(\vec{\bf x})\Big|_{\text{tan-term}}^{N_1}&&= M\, \int_0^{\pi/2} (d \varphi) \,\tan\varphi
\nonumber\\
&&\times
\Big[
\log
\frac{x_1\,\cos\varphi  + x_2 \,\sin\varphi - 1}
{ x_1\,\cos\varphi  + x_2\,\sin\varphi} +
\frac{x_1\,\cos\varphi + x_2\,\sin\varphi}
{ x_1\,\cos\varphi + x_2\,\sin\varphi - 1}
- 1
\Big].
\end{eqnarray}
As above-mentioned, using the replacements (\ref{App-Rep-1}),
the $\eta$-integrated contribution of $f_S^{(\varphi)}(\vec{\bf x})\Big|_{\text{tan-term}}^{N_2}$ can be derived
from $f_S^{(\varphi)}(\vec{\bf x})\Big|_{\text{cot-term}}$. We have
\begin{eqnarray}
\label{IRT-fs-4-2c-App}
f_S^{(\varphi)}(\vec{\bf x})\Big|_{\text{tan-term}}^{N_2}&&=-\, M\, \int_0^{\pi/2} (d \varphi) \,\tan\varphi
\nonumber\\
&&\times
\Big[
\log
\frac{x_2\,\sin\varphi - (1-x_1)\,\cos\varphi}
{ x_1\,\cos\varphi  + x_2\,\sin\varphi} +
\frac{x_1\,\cos\varphi + x_2\,\sin\varphi}
{ x_2\,\sin\varphi - (1-x_1)\,\cos\varphi}
- 1
\Big].
\end{eqnarray}
For our analysis, it is worth to split the contributions from (\ref{IRT-fs-4-2a-App})
and (\ref{IRT-fs-4-2c-App}) into the following contributions:
\begin{eqnarray}
\label{IRT-fs-cot-a-App}
&&f_S^{(\varphi)}(\vec{\bf x})\Big|^{(a)}_{\text{cot-term}}=-\, M\, \int_0^{\pi/2} (d \varphi) \,\cot\varphi
\,
\log
\frac{x_1\,\cos\varphi - (1-x_2)\,\sin\varphi}
{ x_1\,\cos\varphi  + x_2\,\sin\varphi},
\\
\label{IRT-fs-tan-N2-a-App}
&&f_S^{(\varphi)}(\vec{\bf x})\Big|_{\text{tan-term}}^{N_2, (a)}=-\, M\, \int_0^{\pi/2} (d \varphi) \,\tan\varphi
\,
\log
\frac{x_2\,\sin\varphi - (1-x_1)\,\cos\varphi}
{ x_1\,\cos\varphi  + x_2\,\sin\varphi}
\\
\label{IRT-fs-cot-b-App}
&&f_S^{(\varphi)}(\vec{\bf x})\Big|^{(b)}_{\text{cot-term}}=-\, M\, \int_0^{\pi/2} (d \varphi) \,\cot\varphi
\Big[
\frac{x_1\,\cos\varphi + x_2\,\sin\varphi}
{ x_1\,\cos\varphi - (1-x_2)\,\sin\varphi}
- 1
\Big],
\\
\label{IRT-fs-tan-N2-b-App}
&&f_S^{(\varphi)}(\vec{\bf x})\Big|_{\text{tan-term}}^{N_2,\,(b)}=-\, M\, \int_0^{\pi/2} (d \varphi) \,\tan\varphi
\Big[
\frac{x_1\,\cos\varphi + x_2\,\sin\varphi}
{ x_2\,\sin\varphi - (1-x_1)\,\cos\varphi}
- 1
\Big].
\end{eqnarray}

\section{The additional contributions with the quadratic $\tau$-dependence of DRT}
\label{App-3}

The additional contributions with the quadratic $\tau$-dependence of DRT are
\begin{eqnarray}
\label{IRT-fa-4-1-App}
f_A^{(\tau)}(\vec{\bf x})\Big|^{(a)}&&=(i\pi)\, M\,
\int_{0}^{\pi/2}
(d \varphi)
 \frac{A_1(\varphi)}
{ \sqrt{
1- A_1^2(\varphi)} } \theta\big(A_4(\varphi) \big)
\nonumber\\
&&
\times
\Big\{
\theta\big(-A_2(\varphi) \big) +
\theta\big(-A_1(\varphi) \big) +
\theta\big(-A_2^\star(\varphi) \big) \big)
\Big\},
\end{eqnarray}
and
\begin{eqnarray}
\label{IRT-fa-4-11-App}
&&f_A^{(\tau)}(\vec{\bf x})\Big|^{(b)}=(-i\pi)\, M\,
\int_{0}^{\pi/2}
(d \varphi)
 \sqrt{
1-A_1^2(\varphi)}
 \theta\big(A_4(\varphi) \big)
\nonumber\\
&&
\times
\Big\{
\delta\big(A_2(\varphi) \big) +
\delta\big(A_2^\star(\varphi) \big) \big)
\Big\}.
\end{eqnarray}

\section{The additional contributions with the linear $\tau$-dependence of DRT}
\label{App-4}

The additional contributions
with the linear $\tau$-dependence together with the angular $\varphi$-dependence of DRT are
\begin{eqnarray}
\label{fA-sum-1}
f_A^{(\varphi)}(\vec{\bf x})=
f_A^{(\varphi)}(\vec{\bf x})\Big|_{\text{cot-term}} +
f_A^{(\varphi)}(\vec{\bf x})\Big|_{\text{tan-term}}
\end{eqnarray}
where
\begin{eqnarray}
\label{IRT-fa-5-11-App}
&&f_A^{(\varphi)}(\vec{\bf x})\Big|_{\text{cot-term}}=(-i\pi)\, M\,
\int_{0}^{\pi/2}
(d \varphi) \, \cot\varphi \,\theta\big(A_1(\varphi)\big)
\nonumber\\
&&
\times
\Big\{
\theta\big(A_2(\varphi)\big)  -
A_1(\varphi) \delta\big(A_2(\varphi)\big)
\Big\}
\end{eqnarray}
and
\begin{eqnarray}
\label{IRT-fa-5-12-App}
&&f_A^{(\varphi)}(\vec{\bf x})\Big|_{\text{tan-term}}=(i\pi)\, M\,
\int_{0}^{\pi/2}
(d \varphi) \, \tan\varphi \, \theta\big(A_4(\varphi)\big)
\nonumber\\
&&
\times
\Big\{
\theta\big(-A_2^\star(\varphi)\big)  -
A_1(\varphi) \delta\big(A_2^\star(\varphi)\big)
\Big\}.
\end{eqnarray}
We remind that $A_1(\varphi)$, $A_2(\varphi)$, $A_2^\star(\varphi)$, $A_4(\varphi)$ have been defined by (\ref{A-fun-App}).


\end{document}